\documentclass[a4paper,11pt]{article}
    
\usepackage{jheppub} 
  
\usepackage{amsmath,amssymb,xcolor}
\usepackage[english]{babel}
\usepackage{xspace}
\usepackage[normalem]{ulem}

\RequirePackage[T1]{fontenc}

\usepackage{listings}
\newcommand\Rs{R_\text{\sc s}}
\newcommand\BBs{\bar{B}_\text{\sc s}}
\newcommand\BTs{\tilde{B}_\text{\sc s}}

\newcommand{\mathd}{\mathrm{d}}
\newcommand\as{\alpha_{\scriptscriptstyle S}}

\newcommand{\PhiB}{\Phi_{\rm B}}
\newcommand{\Phirad}{\Phi_{\rm rad}}

\newcommand\MCatNLO{{\tt MC@NLO}\xspace}
\newcommand\POWHEG{{\tt POWHEG}\xspace}

\newcommand\KrK{{\tt KrkNLO}\xspace}

\newcommand\tcut{t_{\rm cut}}

\newcommand\NLOPS{{\tt NLOPS}}

\begin{document}

\title{Multiplicative-Accumulative matching of NLO calculations with parton showers}

\author[a,b]{Paolo Nason,}
\author[c,d]{Gavin P.\ Salam}

\affiliation[a]{
  Universit\`a di Milano-Bicocca and INFN, Sezione di Milano-Bicocca,
  Piazza della Scienza 3, 20126 Milano, Italy}
\affiliation[b]{INFN, Sezione di Milano-Bicocca, Piazza della Scienza 3, 20126 Milano, Italy}
\affiliation[c]{Rudolf Peierls Centre for Theoretical Physics, University of Oxford,
  Clarendon Laboratory, Parks Road, Oxford OX1 3PU, United Kingdom}
\affiliation[d]{
  All Souls College, Oxford OX1 4AL, United Kingdom
}

\date{Received: date / Accepted: date}
\preprint{OUTP-21-26P}

\hyphenation{Mon-te}

\abstract{
  We propose a new approach for combining next-to-leading order (NLO)
  and parton shower (PS) calculations so as to obtain three core
  features:
  (a) applicability to general showers, as with the \MCatNLO and
  \POWHEG methods;
  (b) positive-weight events, as with the \KrK and \POWHEG methods;
  and
  (c) all showering attributed to the parton
  shower code, as with the \MCatNLO and \KrK methods.
  This is achieved by using multiplicative matching in phase space
  regions where the shower overestimates the matrix element and
  accumulative (additive) matching in regions where the shower underestimates the
  matrix element, an approach that can be viewed as a combination of
  the \MCatNLO and \KrK methods.
}

\maketitle

\section{Introduction}
Next-to-leading (NLO) order event generators interfaced to Parton Showers (\NLOPS{} from now on) have, in the past
decades, become the state of the art for the simulation of hard collider processes. The \MCatNLO{} algorithm was the first one
to be proposed~\cite{Frixione:2002ik}. Soon after, the \POWHEG{} method was proposed~\cite{Nason:2004rx}
in order to overcome the problem of negative weights inherent to \MCatNLO{}.
Novel methods were proposed later~\cite{Platzer:2012bs,Lonnblad:2012ix,Jadach:2015mza}. Among them, 
the \KrK{} method, like the \POWHEG method, has the characteristic of being free from negative weights, although its applicability
is at the moment restricted to relatively simple processes.

Both the \MCatNLO{} and the \KrK{} methods are intimately intertwined with the parton shower generator that is
adopted. In fact, the two methods provide the
corrections one must apply to the parton shower result in order to achieve
next-to-leading order (NLO) accuracy.
This is at variance with \POWHEG{}, which is largely independent from
the shower generator. In fact, \POWHEG{} takes care of the generation of the hardest event
in such a way that NLO accuracy is preserved, while the PS takes care of the remaining (less hard) radiation.
This feature is convenient, since one can generate events
in the Les Houches format~\cite{Boos:2001cv} and then shower them with any available PS generator that complies
with the Les Houches format specifications.

In view of recent developments aimed at the improvement of the
logarithmic accuracy of the
shower (see Refs.~\cite{Dasgupta:2018nvj,Dasgupta:2020fwr,Hamilton:2020rcu,Karlberg:2021kwr},
\cite{Forshaw:2019ver,Forshaw:2020wrq,Holguin:2020joq}
and \cite{Nagy:2020rmk,Nagy:2020dvz})
it is legitimate to ask
whether \POWHEG{} will maintain the same advantage, i.e. whether minor
modifications to the \POWHEG{} algorithm will be sufficient to
guarantee that, when interfaced to an already NLL accurate shower, NLL
accuracy will be maintained, or whether it will be more
feasible to correct an already NLL accurate shower to achieve also
NLO accuracy, or rather whether both alternatives will be feasible, and
will productively compete among each other.
It is therefore important to explore the full range of options for matching
NLO calculations and parton showers.

In this note, we will address certain limitations of \MCatNLO{} and \KrK{},
and show how they can be overcome.
The  \MCatNLO{} approach is widely applicable, and widely used, but
has the undesirable 
feature of negative weights.\footnote{Recent proposals for the
  reduction of the
  negative-weight fraction include
  Refs.~\cite{Frederix:2020trv,Andersen:2021mvw,Danziger:2021xvr}.}
On the other hand, the \KrK{} method has positive
weights, but it is difficult to extend it to generic processes.
We will show that combining the two methods one could achieve both
positive weights and unrestricted applicability.

The paper is organised as follows, in section~\ref{sec:review} we will compare
the \POWHEG{}, \MCatNLO{} and \KrK{} methods by formulating them in a common language.
We will do this by extending the formulation of the \POWHEG{} and \MCatNLO{} methods of ref.~\cite{Nason:2012pr}
also to the \KrK{} method.
In section~\ref{sec:newmethod} we will present the combined method,
and in section~\ref{sec:variant} we will present some of its possible variants.
Finally, in section~\ref{sec:conc} we present our conclusions.

\section{\POWHEG{}, \MCatNLO{} and \KrK{}}\label{sec:review}
We assume that the phase space with radiation $\Phi$ can be written in terms of
an underlying Born phase space $\PhiB$ and three radiation variables, indicated collectively as
$\Phirad$. We also assume that
the mapping from $\Phi$ to $\PhiB$ is such that in the singular (collinear or soft) limit,
the Born configuration matches the full phase space with 
the collinear pair merged into a single parton or
with the soft particle removed. Such mappings are easy to realise for processes with a single
singular region, while for more complex processes one must separate the real cross section into
contributions having a single singular region. In the following illustrative discussion we ignore these
complications.

We define the following quantity, a function of the underlying Born kinematics,
\begin{equation}\label{eq:barb}
  \BBs(\PhiB)= B_0(\PhiB)+V(\PhiB)+\int \Rs(\Phi) \mathd \Phirad,
\end{equation}
where the full phase space $\Phi$ is defined as function of $\PhiB$ and $\Phirad$,
$B_0$ is the Born cross section, $V$ comprises
  the virtual corrections and, for hadron initiated processes, the collinear counterterms
  integrated at fixed underlying Born,
and $\Rs{}$ is a part of the real cross section that includes all soft and collinear singularities.
In other words, $R-\Rs{}$ is non-singular. Notice that if $\Rs{}$ is taken equal to $R$,
$\BBs{}$ is the inclusive cross section at fixed underlying Born.
We imagine that renormalisation has been carried out, and that the infrared divergences
arising in $V$ and in the $\mathd \Phirad$ integral of $R_s$ are regularised in some way.
Notice also that $\BBs{}$ is finite, since the singularities present
in $V$
cancel those arising integrating the $\Rs$ term.

We also define the Sudakov form factor associated with $\Rs{}$
\begin{align}
  S(t,\PhiB) &= \exp\left[-\int_{t_{\Phi}>t} \frac{\Rs(\Phi)}{B_0(\PhiB)} \mathd \Phirad\right],
\end{align}
where $\Phi$ is defined in terms of the variables $\PhiB$ and $\Phirad$,
and $t$ is some definition of hardness, depending upon the full phase space with radiation.
One may think, for example, that $t$ is the relative transverse momentum of the splitting pair.
If $\Rs$ was taken equal to $R$, the Sudakov form factor would represent the probability for
not radiating anything harder than $t$.

The hardest event cross section can be represented in both \POWHEG{}
and \MCatNLO{} as 
\begin{multline}
  \mathd \sigma =\BBs{}(\PhiB) S(\tcut,\PhiB) \mathd \PhiB 
    +
      \BBs{}(\PhiB)\, S(t_\Phi,\PhiB)
      \times \frac{\Rs(\Phi)}{B_0(\PhiB)} \,\theta(t_\Phi-\tcut) \mathd \Phi
                \,+\\+ \left[R(\Phi)-\Rs{}(\Phi)\right] \mathd \Phi, \label{eq:showerForm1}
\end{multline}
where $\mathd \Phi=\mathd \PhiB\,\mathd \Phirad$, and
$\tcut$ represents a lower limit for radiation, needed to avoid the
Landau-pole singularities.
Events are 
generated with a probability proportional to each term of the cross section formula.
The first two terms are generated with a Monte Carlo technique. In fact, they satisfy the
shower unitarity equation
\begin{equation} \label{eq:unitarity}
S(\tcut,\PhiB) +  \int_{t_\Phi>\tcut} S(t_\Phi,\PhiB)
  \times \frac{\Rs(\Phi)}{B_0(\PhiB)} \mathd \Phirad = 1,
\end{equation}
which follows from the fact that the expression under the integral sign
is an exact differential. In \POWHEG, the generation of the hardness
$t$ is uniform in the Sudakov form factor, and one can generate events
with the standard shower algorithm by equating a random number
with the  Sudakov form factor, and solving for $t$.
If the $t$ value so obtained
is above $t_{\rm cut}$, the radiation kinematics is generated, and the event with the
hardest radiation is fed to a shower generator, which takes care of
adding subsequent (less hard) radiation.

In the case of \MCatNLO{}, the implementation of the first term is generally
more involved. If the shower is ordered in transverse momentum,
the hardest emission is the first, and the radiation mechanism
is the same as in \POWHEG{}, except that it is implemented within the shower
generator, rather than in the NLO program.
In case of an angular ordered shower, large angle soft radiation is generated first,
and the hardest radiation occurs somewhere down the shower.\footnote{Taking $t$ to be the angular
  scale of the shower means that $t$ does not represent a hardness.
  As a result, in the absence of an infrared cutoff,
  the first emission is
  dominated by the infinitely soft region (where $\Rs=R$) rather than
  by the hard region.} It was shown in ref.~\cite{Nason:2004rx}
that, in this case,
by suitable rearrangement of the Sudakov factors for each emission, one reconstructs
the transverse momentum Sudakov form factor in formula~(\ref{eq:showerForm1}).

The last term in
Eq.~(\ref{eq:showerForm1}) is non-singular, and thus is dominated by
hard radiation. It is handled essentially in the same way in \MCatNLO and \POWHEG.

The NLO accuracy of Eq.~(\ref{eq:showerForm1}) can be demonstrated
by computing the expectation value of a generic infrared safe observable $O(\Phi)$ as
follows
\begin{subequations}
  \begin{align}
    \langle O \rangle &=\int \mathd \PhiB \, \BBs{}(\PhiB) \Bigg\{ S(\tcut,\PhiB)\,O(\PhiB) 
                        +\int_{t_\Phi>\tcut} S(t_\Phi,\PhiB)
                        \times \frac{\Rs(\Phi)}{B_0(\PhiB)} O(\Phi) \mathd \Phirad \Bigg\}\,+\nonumber \\
                      &\hspace{16em} \qquad\qquad + \int \left[R(\Phi)-\Rs{}(\Phi)\right] O(\Phi) \mathd \Phi, \\
\intertext{or equivalently,}\label{eq:NLO-accuracy-check-b}
    \langle O \rangle &=\int \mathd \PhiB \BBs{}(\PhiB) \Bigg\{  S(\tcut,\PhiB)\,O(\PhiB)  
                        + \int_{t_\Phi>\tcut} S(t_\Phi,\PhiB)
                        \times \frac{\Rs(\Phi)}{B_0(\PhiB)} O(\PhiB) \mathd \Phirad \,+\nonumber \\
                      &\hspace{12em}+\int_{t_\Phi>\tcut} S(t_\Phi,\PhiB)
                        \times \frac{\Rs(\Phi)}{B_0(\PhiB)}
                        [O(\Phi)-O(\PhiB)] \mathd \Phirad \Bigg\}\,+
                        \nonumber  \\
                      &\hspace{16em}\qquad\qquad
                        + \int \left[R(\Phi)-\Rs{}(\Phi)\right] O(\Phi) \mathd \Phi\,.
  \end{align}
\end{subequations}
In view of the unitarity equation~(\ref{eq:unitarity}), the first two
terms in the curly bracket of equation (\ref{eq:NLO-accuracy-check-b})
collapse into $O(\PhiB)$. Furthermore, the factor $[O(\Phi)-O(\PhiB)]$ in the third
term in the curly bracket suppresses the singular region, so that up to higher order terms
the Sudakov form factor can be omitted, and the lower integration limit can be set to zero
up to a power-suppressed
correction. We thus get
\begin{subequations}
  \begin{align}
    \langle O \rangle &=\int \mathd \PhiB \BBs{}(\PhiB) \Bigg\{O(\PhiB) 
                        +\int \frac{\Rs(\Phi)}{B_0(\PhiB)} [O(\Phi)-O(\PhiB)] \mathd \Phirad \Bigg\}\,+\nonumber  \\
                      &\hspace{12em}+  \int
                        \left[R(\Phi)-\Rs{}(\Phi)\right] O(\Phi) \mathd
                        \Phi  + {\cal O}({\rm NNLO})\,, \\
                      &=\int \mathd \PhiB \left\{ \BBs{}(\PhiB) O(\PhiB) 
                        +\int \Rs(\Phi) [O(\Phi)-O(\PhiB)] \mathd \Phirad \right\}\,+\nonumber  \\
                      &\hspace{12em}+  \int \left[R(\Phi)-\Rs{}(\Phi)\right] O(\Phi) \mathd \Phi  + {\cal O}({\rm NNLO})\,.
  \end{align}
\end{subequations}
Inserting into the above equation the expression for $\BBs$, Eq.~(\ref{eq:barb}), we get
\begin{align}
  \langle O \rangle =\int \mathd \PhiB [B(\PhiB)+V(\PhiB)] O(\PhiB) 
   + \int R(\Phi) O(\Phi) \mathd \Phi+ {\cal O}({\rm NNLO}),
\end{align}
which is the correct NLO expression for the observable.

\POWHEG{} implements Eq.~(\ref{eq:showerForm1}) directly.  $\Rs{}$ is defined as
\begin{equation}
  \Rs{}(\Phi)=F(\Phi) R\,,
\end{equation}
where $F(\Phi)\leq 1$, and approaches 1 as $\Phi$ approaches the singular region,
so that $\Rs{}$ carries all the singularity structure of $R$.

In \MCatNLO{}, $\Rs{}$ is the shower approximation to $R$, typically given by the Born term
times a DGLAP splitting function. Thus, in \MCatNLO{} one computes a Born
configuration using the $\BBs{}$ function, and passes it directly to the Shower Monte Carlo.
The $R-\Rs{}$ terms are generated separately, and fed directly to the shower.
Negative weights can arise at this stage, since there is no guarantee
that  $R-\Rs{}$ is positive.

Using the same language adopted here for \POWHEG{}
and \MCatNLO{}, the key formula for the \KrK{} method can be written
as follows
\begin{equation} \label{eq:KrK}
  \mathd \sigma =\BBs{}(\PhiB) S(\tcut,\PhiB) \mathd \PhiB 
    + \BBs{}(\PhiB) \left\{S(t_\Phi,\PhiB)
    \times \frac{\Rs(\Phi)}{B_0(\PhiB)}\right\} \times \left[ \frac{R(\Phi)}{\Rs(\Phi)} \right] \mathd \Phi\,,
\end{equation}
(with an implicit $\theta(t-\tcut)$ in $d\Phi$).
In the literature dealing with the \KrK{}
method~\cite{Jadach:2017ujd,Jadach:2016qti,Jadach:2016viv,Jadach:2016acv,Jadach:2015mza,Jadach:2015zsq,Jadach:2020xfl},
particular emphasis has been put on the
use of a specific scheme for the parton densities, which  considerably simplifies the expression
for the $\BBs{}$ function.
Here we are instead interested in a simpler aspect of the method,
which is that to generate NLO accurate radiation, it uses a
multiplicative correction, rather than the additive correction of the
\MCatNLO{} method.
The NLO accuracy of the \KrK{}
formula can be simply demonstrated by showing that \KrK{} is equivalent to \MCatNLO{} at the NLO level.
In fact, we can rewrite formula~(\ref{eq:KrK}) as
\begin{multline} \label{KrK}
  \mathd \sigma =\BBs{}(\PhiB) S(\tcut,\PhiB) \mathd \PhiB 
  +
  \BBs{}(\PhiB) S(t_\Phi,\PhiB)
    \times \frac{\Rs(\Phi)}{B_0(\PhiB)} \times \left[ \frac{R(\Phi)}{\Rs(\Phi)}-1 \right] \mathd \Phi
    +\\+
    \BBs{}(\PhiB) S(t_\Phi,\PhiB) \frac{\Rs(\Phi)}{B_0(\PhiB)}   \mathd \Phi\,.
\end{multline}
The middle term is now insensitive to the soft region, because the factor in square brackets vanishes there,
so we can drop the Sudakov form factor and, neglecting terms of NNLO size, set $\BBs{}/B_0=1$. By
doing this we recover exactly Eq.~(\ref{eq:showerForm1}).

The \KrK{} method leads to positive weighted events. On the other hand, unlike \MCatNLO{} and \POWHEG{},
its cross section at fixed underlying Born does not exactly match the
corresponding fixed order result, but differs from it by NNLO terms.\footnote{Although some authors consider this to
be an undesirable feature, it does not constitute a real problem, since choices for
uncontrolled NNLO terms are made, for example, when choosing the scales, and there is no preferred
way to define an NLO result as far as the neglected NNLO terms are concerned.}
The \KrK{} method, however, generates weighted events, so, the unweighting efficiency may constitute a problem
if one wants to apply the method to generic processes without having to do process-by-process adjustments.
As a related problem, the shower generator may not cover the
full radiation phase space. This is the same as saying that $\Rs{}$ can become zero in certain regions, in which case
the method is not applicable. For these
reasons, it seems difficult to apply the method to generic processes in automated framework, something
that has been available in \POWHEG{} and \MCatNLO{} since a long time.

\section{The new method}\label{sec:newmethod}
By comparing the \MCatNLO{} and \KrK{} method using a common language,
it becomes clear that the two methods 
have much in common, and can in fact be merged in such a way that the \KrK{} positivity is maintained, unweighted
events can be generated on the fly, and no issues arise from the
limited coverage of the phase space by the parton shower code.
The merged method is defined by the following formula
\begin{equation}\label{eq:hybrid}
  \mathd \sigma = \BBs{}(\PhiB) \, S(t_\Phi,\PhiB)
                  \times \frac{\Rs(\Phi)}{B_0(\PhiB)}  
                 \times \left\{1+ \frac{R-\Rs{}}{\Rs} \theta(\Rs-R)\right\} \mathd \Phi 
                +\theta(R-\Rs) \left[R-\Rs\right] \mathd \Phi\,.
\end{equation}
It is easy to see by inspection that this formula has the same NLO accuracy as \MCatNLO{}. In fact, the $\theta$ function
in the first term of eq.~(\ref{eq:hybrid}) is regular,
and as such the Sudakov form factor multiplying it and the $\BBs{}/B_0$ ratio
can both be set to 1 up to NNLO corrections. Having done this, the contribution proportional
to the $\theta$ function in the first term becomes identical to the
last term, except
for the theta function, which has the opposite sign in the argument. The two terms thus combine, and the theta function
disappears, yielding the \MCatNLO{} formula.

The method of Eq.~(\ref{eq:hybrid}) is immediately seen to have
positive weights, since the second term 
in eq.~(\ref{eq:hybrid}) is forced to be positive by the
theta function, while the factor in the curly bracket of the 
first term
is forced
by the theta function to be less than 1. Notice that regions of phase
space not populated by the Monte Carlo only contribute to the additive
term in the square bracket, since they have $\Rs=0$.

Let us consider how the method of Eq.~(\ref{eq:hybrid}) might be
implemented in a scenario where it is the parton shower that is
responsible for generating $\Rs$.\footnote{Alternatively, one might
  also implement the approach in a \POWHEG-style scheme, where the NLO
  program takes responsibility for the first emission, and $\Rs$ is not required
  to be smaller than $R$.}
We first consider the case of a parton shower ordered in a genuine
hardness variable (for example, transverse momentum).
As in the \MCatNLO approach, the NLO program is responsible for
generating a sample of Born events, with a distribution corresponding
to the correct $\BBs{}(\PhiB)\mathd\PhiB$ weight.
The event is communicated to the parton shower, which generates a
first emission, i.e.\ accounting for a remaining factor
$S(t_\Phi,\PhiB) \frac{\Rs(\Phi)}{B_0(\PhiB)} \mathd\Phi/\mathd\PhiB$
in Eq.~(\ref{eq:hybrid}).
Next, the parton shower communicates the event, with its first
emission, back to the NLO program, which evaluates the contents of the
curly bracket in Eq.~(\ref{eq:hybrid}).
The result of that evaluation, which is bounded to be between zero and
one, is to be used as an acceptance probability for the event.
If the event is accepted, the parton shower then continues from that
point, generating the remaining emissions.
In addition, the NLO program is responsible for generating a second
(positive definite) sample of events, corresponding to the
$\theta(R - \Rs$) term of Eq.~(\ref{eq:hybrid}), covering the part of
the real phase space where the shower $\Rs$ underestimates the true
real matrix element.
These events are passed to the parton shower code for normal
showering.

The above scheme requires a more connected workflow between the parton
shower and the NLO code than either of the \MCatNLO or \POWHEG
approaches.
Recall that those approaches pass a Les Houches Event file to the
shower code, and then let the shower take over from there.
In contrast, the scheme outlined above requires additional action from
the NLO code \emph{after} the first parton shower emission.
However, this is simply a technical consideration, which in our view
is a small price to pay for an approach that, like the \MCatNLO
scheme, leaves responsibility for the first emission with the parton
shower, while eliminating negative weights.

A further remark concerns angular-ordered parton showers.
For such showers, the application of the acceptance probability
(factor in curly brackets in Eq.~(\ref{eq:hybrid})) after the first
emission would not be correct.
Instead, one might envisage an approach in which one runs the complete
shower, uses a jet algorithm to map the full event to the real phase
space and then applies the acceptance probability.

\section{Variants of the method}\label{sec:variant}

Eq.~(\ref{eq:hybrid}) can be viewed as a special case of a family of
approaches, parameterised by a constant $c \ge 1$,
\begin{multline}\label{eq:hybridvar}
  \mathd \sigma = \BBs{}(\PhiB) \, S(t_\Phi,\PhiB)
                  \times \frac{c\,\Rs(\Phi)}{B_0(\PhiB)} 
                 \times \left\{1+ \frac{R-c\,\Rs{}}{c\,\Rs}
                  \theta(c\,\Rs-R)\right\} \mathd \Phi \,+ \\
                +\theta(R-c\,\Rs) \left[R-c\,\Rs\right] \mathd \Phi\,.
\end{multline}
This formula was obtained by replacing $\Rs \to c\,\Rs$ in all
occurrences where it appears in Eq.~(\ref{eq:hybrid}), except for the
Sudakov form factor.
First of all, we should convince ourselves
that this formula has the correct behaviour near the soft limit, and
that it is NLO accurate. This is seen immediately
by writing~(\ref{eq:hybridvar}) as
\begin{multline}
  \mathd \sigma = \BBs{}(\PhiB) \, S(t_\Phi,\PhiB)
                  \times \frac{c\,\Rs(\Phi)}{B_0(\PhiB)}
                  \times \left\{1+ \frac{R-c\,\Rs{}}{c\,\Rs}
                  \right  \} \mathd \Phi\,,  \\
                + \theta(R-c\,\Rs) (R-c\,\Rs)
                  \left[1-\frac{\BBs{}(\PhiB)}{B_0(\PhiB)} \, S(t_\Phi,\PhiB)
                  \right]  \mathd \Phi\,.
                  \label{eq:hybridvar1}
\end{multline}
The first line is equal to the \KrK{} formula, eq.~(\ref{eq:KrK})
(leaving aside the $\tcut$, for simplicity).
In the second line, the factor in square brackets is of order $\as$
and it multiplies a term  dominated by the hard region (and thus also of order $\as$), since
\begin{displaymath}
  \theta(R-c\,\Rs) (R-c\,\Rs) \le  \theta(R-\Rs)(R-\Rs)\,.
\end{displaymath}
Accordingly, the second line is of NNLO order, and we can conclude
that formula~(\ref{eq:hybridvar}) is equivalent to the \KrK{} formula
up to NNLO corrections.
For a shower where $\Rs / R$ is always larger than some value $r$,
taking $c \ge 1/r$ causes Eq.~(\ref{eq:hybridvar}) to reduces to the
\KrK approach over all of phase space.

The implementation of formula~(\ref{eq:hybridvar}) goes as follows. The soft
events are oversampled by a factor of $c$, and accepted with a probability
proportional to the expression in the curly bracket, while the hard events
are generated in a standard way.

There are several reasons why it may be of use to have such a family of
approaches parameterised by $c$.
One is that it is useful to have a parameter to help gauge systematic
uncertainties associated with terms that are beyond the accuracy of
the method. In fact, the $c$ parameter also gauges the amount of radiation
that is multiplied by the inclusive $K$-factor, with respect to the amount
that is added in as hard radiation. In this sense, it would play a similar
role to the {\tt hdamp} parameter in
\POWHEG{}~\cite{Nason:2004rx,Alioli:2008tz}.
Another reason is that the $\theta$-functions in
Eqs.~(\ref{eq:hybrid}), (\ref{eq:hybridvar}) can, conceivably, introduce
non-smoothness in phase-space coverage even when the underlying $\Rs$
and $R$ functions are smooth.
Sampling over a range of $c$ values would allow one to address that
issue.\footnote{It is not clear to us that this would be a genuine
  problem.
  A worse issue is potentially present in the
  \MCatNLO approach (and also in Eqs.~(\ref{eq:hybrid}) and
  (\ref{eq:hybridvar})) if a shower has a discontinuity in the
  distribution of the hardness variable.
  We are not aware of this having caused significant problems in
  practice, possibly because subsequent showering smoothens any
  discontinuities.
  Were it to be a problem, one solution could be to sample over a
  range of shower starting scales.
}

So far, the variants that we have discussed involve the rejection of
events.
For showers with an ordering variable that involve a genuine hardness
scale, such as transverse momentum, it is also possible to envisage a
variant where instead of rejecting the event with probability
$\frac{\Rs-R}{\Rs} \theta(\Rs - R)$, one rejects the shower's first
emission with that same probability.
If that first emission is rejected, at a value of the ordering
variable that we label $t_1$, the shower then continues from that
scale $t_1$, based on an event without the first emission.
As the shower continues, each time the shower again attempts to create
a new first emission, that emission continues to be rejected with
probability $\frac{\Rs-R}{\Rs} \theta(\Rs - R)$.
Once the shower has generated a first emission that is accepted, the
shower continues as normal.
This ensures that the shower (with first emission rejection) remains
unitary.\footnote{As a matching algorithm to ensure the correct matrix
  element for the first emission, it bears strong similarities to the
  algorithm of Refs.~\cite{Bengtsson:1986et,Seymour:1994df}.
  That algorithm works within the assumption that one can find some
  constant $c$ such $c \Rs > R$ over the full phase space, evading the
  need for an additive term.}
The Born normalisation factor that multiplies the shower generation
then needs to be modified to read
\begin{equation}\label{eq:tildeb}
  \BTs(\PhiB)= B_0(\PhiB)+V(\PhiB)+\int \min[R(\Phi),\Rs(\Phi)] \mathd \Phirad\,.
\end{equation}
With this variant, the distribution of the hardest radiation is given by
\begin{equation}
  \mathd \sigma = \BTs{}(\PhiB) \, \tilde S(t_\Phi,\PhiB)
                  \times \frac{\min(R(\Phi),\Rs(\Phi))}{B_0(\PhiB)}  
                 \mathd \Phi 
                +\theta(R-\Rs) \left[R-\Rs\right] \mathd \Phi\,,
\end{equation}
where $\tilde S(t_\Phi,\PhiB)$ is the Sudakov form factor that is
effectively obtained as a result of the emission rejection procedure,
\begin{align}
  \tilde S(t,\PhiB) = \exp\left[-\int_{t_{\Phi}>t}
  \frac{\min[R(\Phi),\Rs(\Phi)]}{B_0(\PhiB)} \mathd \Phirad\right].
\end{align}

\section{Conclusion}\label{sec:conc}

In this article we have proposed an algorithm for NLOPS generators
that, like the \MCatNLO approach, relies on the parton shower code for
driving the showering steps, but avoids the negative-weight events
that complicate the practical use of the \MCatNLO approach.
The approach can be viewed as a hybrid version of the \MCatNLO{} and
\KrK{} methods.
The method should be straightforward to implement, requiring no more
information than is already used for \MCatNLO codes, though it should
be noted that it will probably require a closer integration of the
showering and fixed-order codes.
Following the tradition of associating an acronym to
NLOPS methods, a suitable one for this method could be {\tt MAcNLOPS},
reflecting the combination of \underline{M}ultiplicative and
\underline{Ac}cumulative (additive) steps.

\appendix

\section*{Acknowledgements}
P.N.\ wishes to thank Stanislaw Jadach, Andrzej Si\'odmok, Silvia Ferrario
Ravasio, James Whitehead and Giulia Zanderighi for helpful conversations.
G.P.S.\ wishes to thank his collaborators on the PanScales project
(Melissa van Beekveld,
Mrinal Dasgupta,
Fr\'ed\'eric Dreyer,
Basem El-Menoufi,
Silvia Ferrario Ravasio,
Keith Hamilton,
Alexander Karlberg,
Rok Medves,
Pier Monni,
Ludovic Scyboz,
Alba Soto Ontoso,
Gr\'egory Soyez
and Rob Verheyen) for discussions on these and related
topics.
P.N.\ acknowledges support from Fondazione Cariplo and Regione
Lombardia, grant 2017-2070, and from INFN.
G.P.S.\ acknowledges support from a Royal Society Research
Professorship (RP$\backslash$R1$\backslash$180112),
the European Research Council (ERC) under the European Union’s Horizon
2020 research and innovation programme (grant agreement No.\ 788223,
PanScales)
and the Science and Technology Facilities Council (STFC) under
grant ST/T000864/1.

\bibliography{paper}

\end{document}